\documentclass[preprint,a4paper]{aastex}
\usepackage{graphics,epsf}
\usepackage{amsmath}                
\usepackage{amsfonts}               
\usepackage{amssymb}                
\usepackage{epsfig}                 
\usepackage{rotating}

\usepackage{subfigure}

\newcommand{\kms}{{~\rm km\; s^{-1}}}

\newcommand{\cm}{{~\rm cm}}
\newcommand{\s}{{~\rm s}}
\newcommand{\km}{{~\rm km}}
\newcommand{\g}{{~\rm g}}
\newcommand{\K}{{~\rm K}}
\newcommand{\erg}{{~\rm erg}}
\newcommand{\yr}{{~\rm yr}}
\newcommand{\Myr}{{~\rm Myr}}

\newcommand{\kpc}{{~\rm kpc}}

\newcommand{\keV}{{~\rm keV}}

\def \apj{ApJ}
\def \aap{A\&A}

\def \mnras{MNRAS}
\def \apjl{ApJ Lett.}
\def \apjs{ApJ Suppl. Ser.}
\def \nat{Nature}

\begin{document}

\title{RESCUING THE INTRACLUSTER MEDIUM OF NGC 5813}

\author{Noam Soker\altaffilmark{1}, Shlomi Hillel\altaffilmark{1},  and Assaf
Sternberg\altaffilmark{2}}

\altaffiltext{1}{Department of Physics, Technion -- Israel
Institute of Technology, Haifa 32000, Israel;
soker@physics.technion.ac.il; shlomihi@tx.technion.ac.il}

\altaffiltext{2}{Geneva Observatory, University of Geneva, Chemin des Maillettes 51, CH-1290 Versoix, Switzerland; assaf.sternberg@unige.ch}



\begin{abstract}
We use recent X-ray observations of the intracluster medium (ICM)
of the galaxy group NGC~5813 to confront theoretical studies of
ICM thermal evolution with the newly derived ICM properties. We
argue that the ICM of the cooling flow galaxy group NGC~5813 is
more likely to be heated by mixing of post-shock jets' gas
residing in hot bubbles with the ICM, than by shocks or
turbulent-heating. Shocks thermalize only a small fraction of their energy in
the inner regions of the cooling flow; in order to adequately heat
the inner part of the ICM, they would overheat the outer regions by
a large factor, leading to its ejection from the group.
Heating by mixing, that was found to be much more efficient than turbulent-heating and
shocks-heating, hence, rescues the outer ICM of NGC~5813 from its
predestined fate according to cooling flow feedback scenarios that
are based on heating by shocks.
\end{abstract}

{\bf Keywords: }  galaxies: active – galaxies: clusters: general –
galaxies: groups: individual (NGC 5813) - galaxies: jets

\section{INTRODUCTION}
\label{s-introduction}

A negative feedback mechanism driven by active galactic nucleus
(AGN) jets that inflate X-ray bubbles (X-ray deficient cavities)
determines the thermal evolution of the intra-cluster medium (ICM)
in the inner regions of cooling flow (CF) clusters and groups of
galaxies (e.g., \citealt{Pfrommer2013, Peruchoetal2014}). In the
feedback cycle the ICM feeds the AGN, and the AGN launches jets
that heat the ICM.

A consensus is emerging that the feeding of the AGN is through
cold clumps, in what is termed the \emph{cold feedback mechanism}
\citep{Pizzolato2005}. Observations and theoretical studies in
recent years have put the cold feedback mechanism on a very solid
ground (e.g., \citealt{Revaz2008, Pope2009, Pizzolato2010,
Edge2010, Wilman2011, Nesvadba2011, Cavagnolo2011, Gaspari2012a,
Gaspari2012b, McCourt2012, Sharma2012, Farage2012, Waghetal2014,
BanerjeeSharma2014, McNamaraetal2014, VoitDonahue2015,
Voitetal2015, Lietal2015, Prasadetal2015, Russelletal2015,
Tremblayetal2015, Fogartyetal2015}).

The processes by which jets heat the ICM, on the other hand, is
still in dispute. Based on observations of the galaxy group
NGC~5813 \cite{Randalletal2011} and \cite{Randalletal2015} argue
that shocks heat the ICM; hereafter shocks-heating process. This
was put into question by \cite{Sokeretal2013}, and is further
studied in the present paper. Based on deep X-ray observation of
the Perseus and Virgo cooling flow clusters,
\cite{Zhuravlevaetal2014} argue that the main heating process is
via dissipation of ICM turbulence; hereafter turbulent-heating
process. The third heating process is the mixing of cooling ICM
with hot shocked jets' gas that reside in the hot bubbles;
hereafter mixing-heating process. \cite{GilkisSoker2012} and
\cite{HillelSoker2014} conducted numerical simulations and showed
that the mixing-heating process is much more efficient than the
shocks-heating process. In a recent paper \cite{HillelSoker2016}
showed that the mixing-heating is also more efficient than the
turbulent-heating when turbulence is driven by AGN activity. The
turbulence, as observed in some clusters (e.g.,
\citealt{Zhuravlevaetal2015, AndersonSunyaev2015}), is a
consequence of the same vortices that cause mixing.

To reach the conclusion that mixing-heating is the dominant
process, it is mandatory to inflate bubbles by jets in a
self-consistent manner. For example, the vortices that are induced
inside and outside the bubbles must be taken into account as they
play major roles in the formation of bubbles, their evolution, and
their interaction with the ICM, (e.g. \citealt{Omma2004,
Heinz2005, Roediger2007, Sternberg2008b, GilkisSoker2012,
Walgetal2013}). To inflate bubbles that resemble observed bubbles,
our group has employed either slow (sub-relativistic) massive
wide (SMW) jets (bipolar outflows), e.g., \citep{Sternberg2007},
or precessing jets, e.g., \citep{Sternberg2008a}; see also
\cite{Falceta-Goncalves2010}. SMW jets are supported by
observations (e.g., \citealt{Moe2009, Dunn2010, Aravetal2013}).
Both types of inflated bubbles will be described in the present
study.

In a recent paper \cite{Randalletal2015} presented new
observations of NGC~5813. Such detailed observations serve as a
unique opportunity for us to confront our findings, regarding
heating the ICM with jet-inflated bubbles, against observations.
We here confront the new observations with two types of previously
conducted numerical simulations. In section \ref{sec:flow} we
present a new study of the implications on the heating process of
large scale flows that are induced by the jet-inflated bubbles. In
section \ref{sec:post} we show that a careful treatment is
required when studying the shocks that are induced by jet-inflated
bubbles. To our best knowledge, this is the first study that point
to the influence of sound waves on the derived shock properties.
But we first present in section \ref{sec:weak} a new analysis of
the effect on many repeated shocks on the outskirts of the ICM in
NGC~5813. In section \ref{s-summary} we summarize our findings,
and conclude that although shocks play some role, mixing-heating
is likely to play a much more significant role in the heating
process.

\section{IMPLICATIONS FOR THE OUTER ICM}
\label{sec:weak}

The inefficiency of the shocks-heating process was discussed in
\cite{Sokeretal2013}. To excite a shock with a Mach number of
${\mathcal{M}_{\rm s}}$  at radius $r$, the energy supplied by the
AGN directly to the ICM, using the spherically symmetric Sedov solution for the shock, should be
\begin{equation}
 E_{\rm shock} \approx  \left( \frac{\mathcal{M}_{\rm s}}{1.1} \right)^2
 E_{\rm th} (r),
 \label{eq:Mach}
\end{equation}
where $E_{\rm th}(r)$ is the thermal energy of the ICM inside
radius $r$. According to \cite{Randalletal2015}, to heat the
cooling flow region of NGC~5813 inside $30 \kpc$, about 140 shocks
with a Mach number of $\mathcal{M}_a \approx 1.2$ are required
during one cooling time. By equation (\ref{eq:Mach}), to reach this Mach number each shock is excited with an energy of $E_{\rm shock} \approx (1.2/1.1)^2 E_{\rm in} =1.2 E_{\rm in}$, where we define $E_{\rm in} \equiv E_{\rm th} (r < 30 \kpc)$. Each shock supplies an energy of $\approx E_{\rm in}/140=0.01 E_{\rm in}$. Removing this dissipated energy in the inner region (up to $30 \kpc$), the energy that flows to the outer ICM regions within one cooling time at $r=30 \kpc$ is $E_{\rm out} \approx 140 (1.2-0.01) E_{\rm in} \approx 160 E_{\rm in}$.

The ICM mass residing out at $r>30 \kpc$ can be crudely estimated
from the new density profile given by \cite{Randalletal2015}. The
electron density profile from $r=22 \kpc$ to $r=35 \kpc$, the
largest radius in their profile, can be fitted with
\begin{equation}
 n_e(r)= 2.4 \times 10^{-3} \left( \frac{r}{30 \kpc}
 \right)^{-2.8} \cm^{-3}.
 \label{eq:ne}
\end{equation}
Extrapolating this fit to large distances of hundreds of $\kpc$,
we can estimate the mass in the outer region to be $M_{\rm out}
\approx 10^{11} M_\odot \approx 2 M_{\rm in}$.
{{{{ Based on ROSAT observations of poor groups of galaxies \cite{Mulchaeyetal1996} find the diffuse X-ray gas in a typical poor group to extend up to $\sim 300 \kpc$, and have a shallower density profile.
If we take a shallower density profile of
\begin{equation}
n_{e-s}(r)= 2.4 \times 10^{-3} \left( \frac{r}{30 \kpc}
 \right)^{-1.5} \cm^{-3},
  \label{eq:ne2}
\end{equation}
 (like in eq. \ref{eq:rhoICM} below), the gas mass in the outer region to a distance of $300 \kpc$ is $M_{{\rm out}-s} \approx 5 \times 10^{11} M_\odot \approx 10 M_{\rm in}$}}}}

\cite{Randalletal2015} do mention that the energy carried by the
bubble is transferred out by the buoyant bubbles. But the shocks
turn into sound wave that dissipate their energy after a long time
in the ICM. By the simple estimate made above the temperature of
the outer gas should be
\begin{equation}
T_{\rm out} \approx 10
 \left( \frac{T_{\rm in}}{0.7 \keV} \right)
 \left( \frac{E_{\rm out}}{150E_{\rm in}} \right)
 \left( \frac{M_{\rm out}}{10 M_{\rm in}} \right)^{-1}  \keV.
 \label{eq:tout}
\end{equation}
This is a temperature larger than the virial temperature of the group.

It seems that for the shocks-heating process to be effective in
the inner region, the outer ICM would escape in a time much
shorter than the cooling time, hence destroying the ICM observed
structure. Clearly, a much more efficient heating mechanism than
shocks is required to rescue the ICM of NGC~5813. Mixing-heating
is this process \citep{HillelSoker2014, HillelSoker2016}.

\section{LARGE SCALE FLOW}
\label{sec:flow}

In the shocks-heating scenario presented by \cite{Randalletal2011}
and \cite{Randalletal2015} tens of jet-launching episodes along
the same axis heat the ICM.  They do not consider any flow that
might develop in the ICM. We here show that a large-scale
meridional flow develops as a result of multi-jet activity cycles.
For that we present results from 3D numerical simulations we have
carried out recently \citep{HillelSoker2016}. We here briefly
present the main characteristics of the numerical set up; more
details can be found in that paper.

We use the {\sc pluto} code \citep{Mignone2007} for the
hydrodynamic simulations in a three-dimensional Cartesian grid
with adaptive mesh refinement (AMR). The computational grid is in
the octant where the three coordinates $x$, $y$ and $z$ are
positive. At the $x = 0$, $y = 0$ and $z = 0$ planes we apply
reflective boundary conditions. The $z$ coordinate is chosen along
the initial axis of the jets. The computational grid spans the
cube $0 \leq x, y, z \leq 50 \kpc$.

At the boundary $z = 0$ we inject into the grid a jet with a
half-opening angle of $\theta_{\rm j} = 70^\circ$
\citep{Sternberg2007}. The jet is injected during each active
episode lasting $10 \Myr$, and when the jet is turned off for $10
\Myr$, reflective boundary conditions apply in the whole $z = 0$
plane. The initial jet velocity is $v_{\rm j} = 8200 \kms$
corresponding to a Mach number of about $10$ relative to the ICM.
The power of the two jets together (we simulate only one jet) is
$\dot E_{2{\rm j}} = 2 \times 10^{45} \erg \s^{-1}$ (half of it in
each direction), and the mass deposition rate is $\dot{M}_{2{\rm
j}} = 94 M_{\odot}~\yr^{-1}$.

{{{{ The simulations presented here assume jets starting with wide opening angles. As stated in section \ref{s-introduction}, there are observations of slow massive wide outflows from AGNs. As well, the effect of rapidly precessing jets in inflating bubbles is very similar to that of wide jets \citep{Sternberg2008a}, but it is simpler to simulate wide jets (for precessing jets see next section). The main issue is to inflate bubbles, a process that involves the formation of vortices. Although magnetic fields are present in jets, as inferred from the radio emission, we here assume that they do not play a dynamical role. As we heat by mixing, we attribute no role to heat conduction on large scales (only on very small scales as mixing brings the ICM and the hot bubble gas very close to each other). }}}}

The simulation begins with an isothermal box of gas at an initial
temperature of $T_{\rm ICM} (0) = 3 \times 10^7 \K$ with a density
profile of (e.g., \citealt{VernaleoReynolds2006})
\begin{equation}
\rho_{\rm ICM}(r) = \frac{\rho_0}{\left[ 1 + \left( r / a \right)
^ 2 \right] ^ {3 / 4}},
\label{eq:rhoICM}
\end{equation}
with $a = 100 \kpc$ and $\rho_0 = 10^{-25} \g \cm^{-3}$. A gravity
field is added to maintain an initial hydrostatic equilibrium, and
is kept constant in time. We include radiative cooling in the
simulations, where the tabulated cooling function is taken from
Table 6 in \cite{SutherlandDopita1993}.

To reveal the large scale flow we follow an artificial flow
quantity called 'tracer.' The tracer is frozen-in to the flow, and
hence tells us on the spreading with time of gas starting in a
certain volume. A tracer's initial value is set to $\xi (0) = 1$
in a certain volume and $\xi (0) = 0$ elsewhere. As the traced gas
mixes with the ICM or the jet's material, its value drops to $0 <
\xi(t) < 1$. We choose to trace the gas starting inside a torus
around the $z$ axis. The radius of the cross section of the torus
is $2.5 \kpc$, and it is centered at $(y,z)=(20,15)\kpc$; or more
generally $(\sqrt{x^2+y^2},z)=(20,15)\kpc$.

\begin{figure*}
\centering
\subfigure{\includegraphics[width=0.45\textwidth]{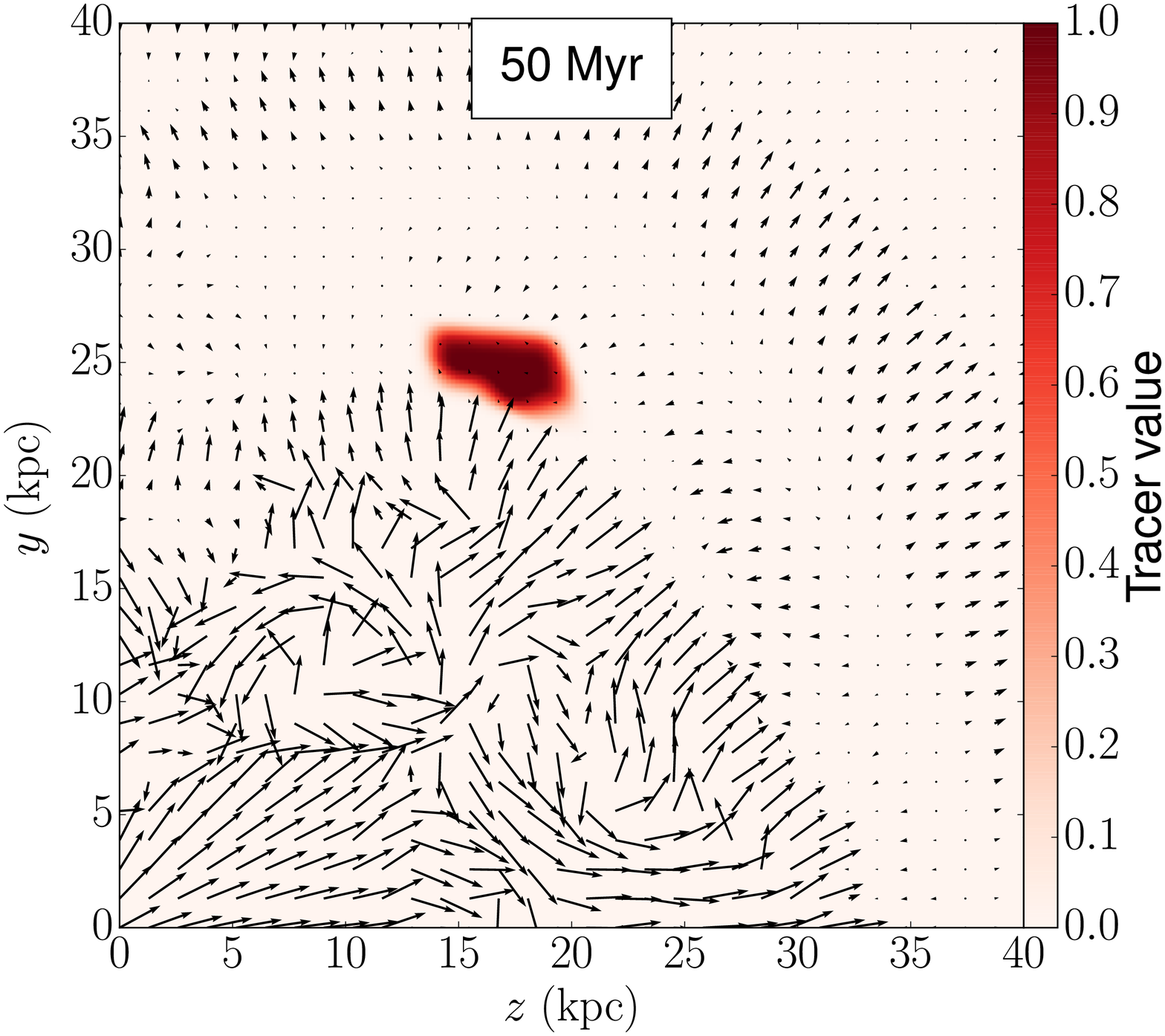}}
\subfigure{\includegraphics[width=0.45\textwidth]{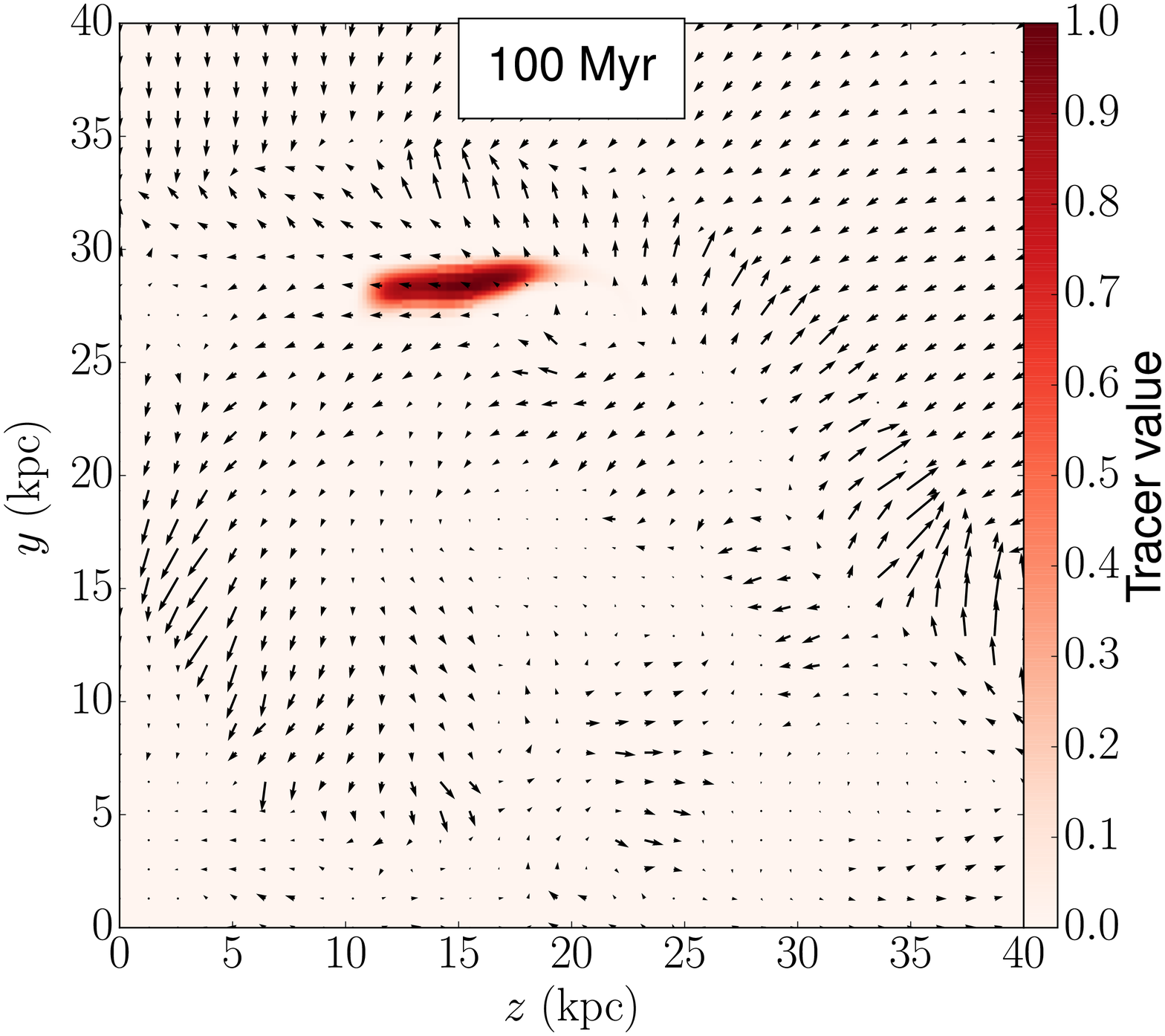}}\\
\subfigure{\includegraphics[width=0.45\textwidth]{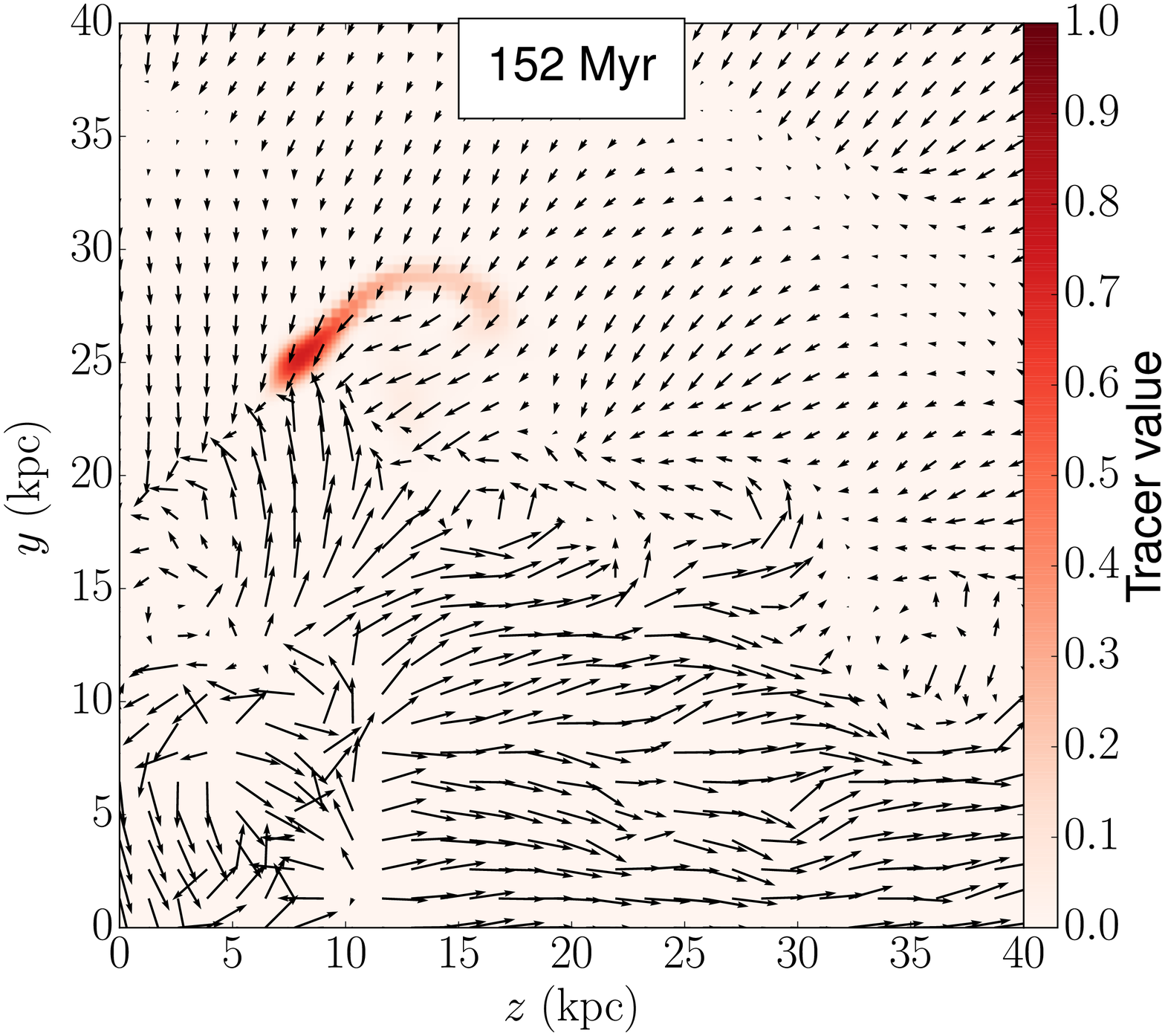}}
\subfigure{\includegraphics[width=0.45\textwidth]{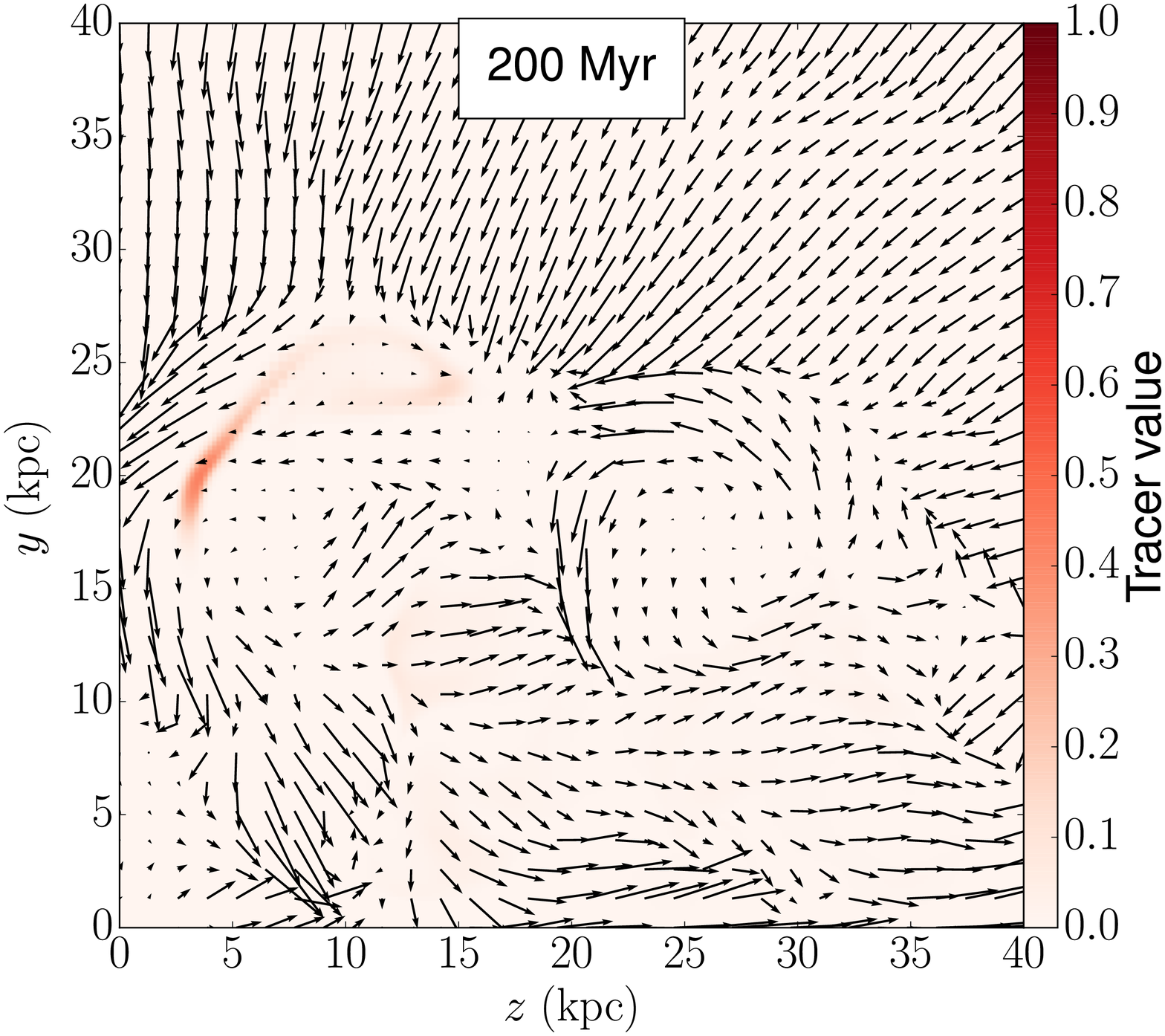}}\\
\subfigure{\includegraphics[width=0.45\textwidth]{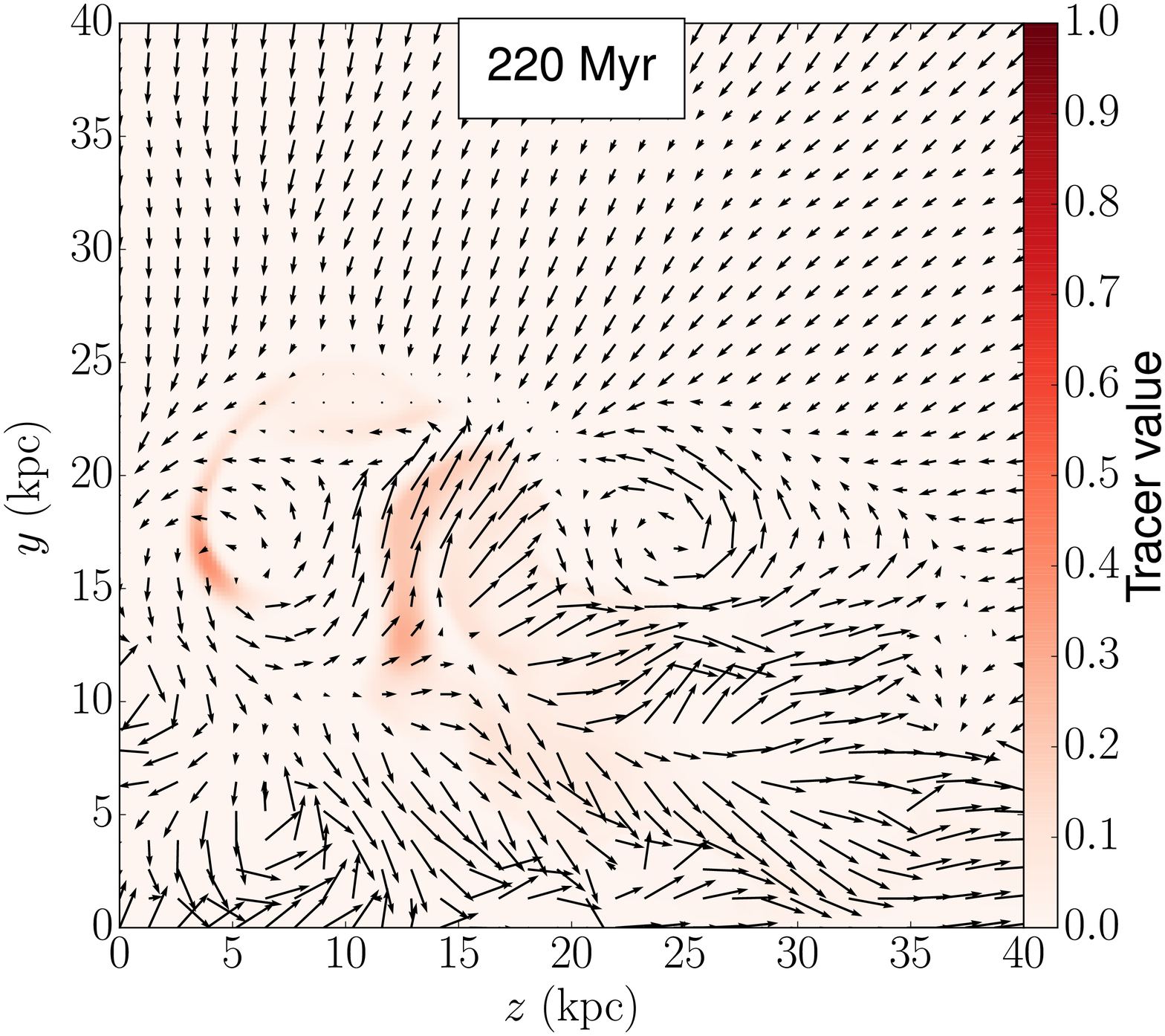}}
\subfigure{\includegraphics[width=0.45\textwidth]{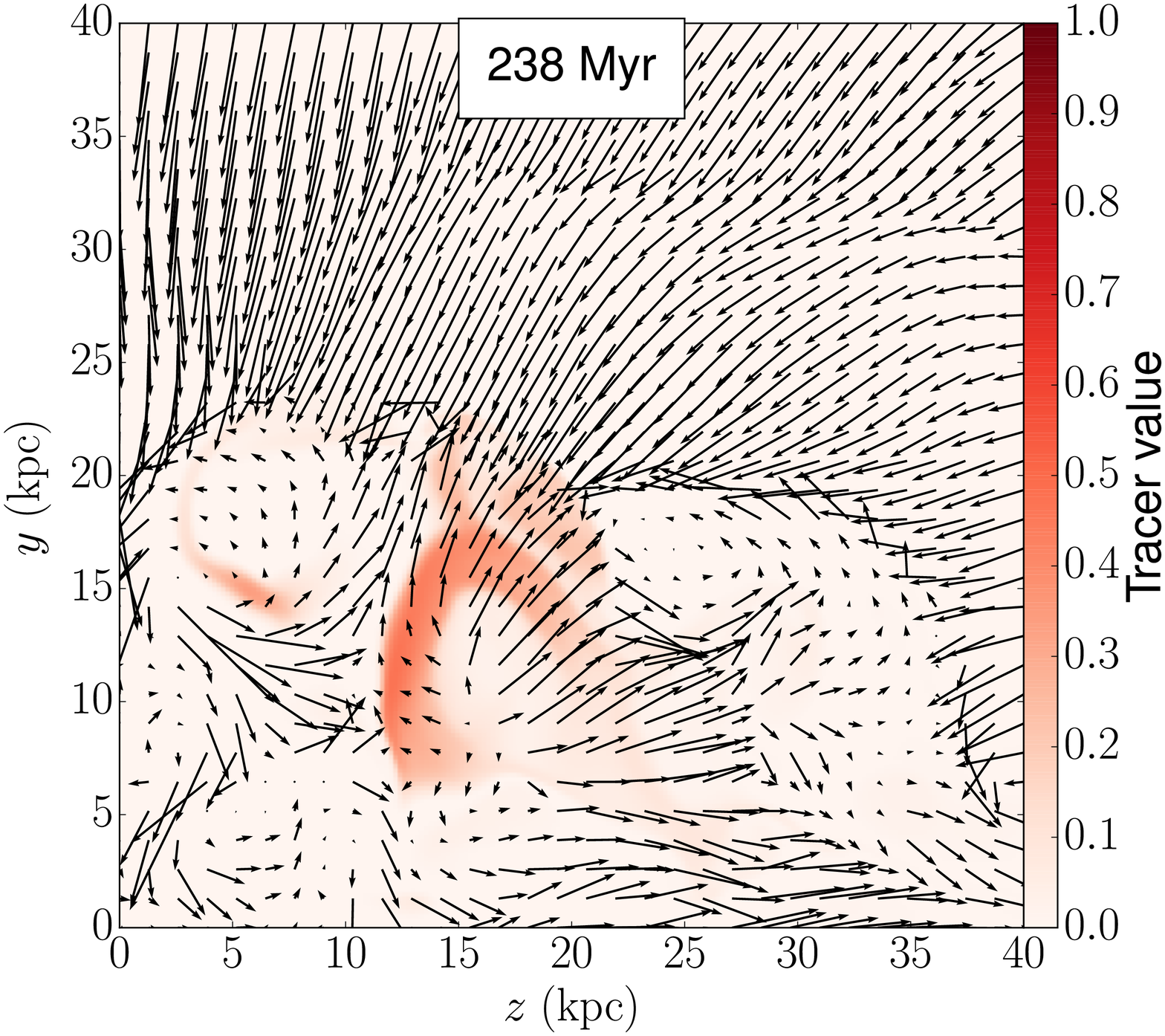}}
\caption{Evolution with time of a traced gas simulated by \cite{HillelSoker2016}. At $t=0$ the traced gas was contained in a torus
whose axis is the $z$-axis and whose cross section is a circle
centered at $(\sqrt{x^2+y^2},z)=(20,15)\kpc$ with a radius of
$r=2.5\kpc$. Color coding is of the tracer value $\xi$. The jets'
axis is along the horizontal $z$-axis in the panels. In the three
panels on the left the arrow lengths are linear with velocity, up
to a maximum value of $v_m = 400 \kms$. Faster regions have an
arrow length corresponding to $v_m$. The arrows in the right
panels show the mass flux $\phi=v \rho$. A length of $1 \kpc$ on
the map corresponds to $\phi = 2.3 \times 10^{-23} \km \s^{-1} \g
\cm^{-3}$. }
 \label{figure:Hillel}
\end{figure*}

In Fig.~\ref{figure:Hillel} we present the evolution of the tracer
at six times in the meridional plane $(y,z)$. Note that the jets'
axis is along the $z$ axis, which is the horizontal axis in the
figure. The evolution of other quantities in the $(x,z)$
meridional plane is presented by \cite{HillelSoker2016}. In the
panels on the left column the arrows represent flow velocity, with
length linear with velocity up to an upper limit of $v_m = 400
\kms$. Faster regions are presented with an arrow length
corresponding to $v_m$. In the panels on the right column the
arrows represent mass flux $\phi = v \rho$.

There are several prominent properties of the flow that are
relevant to the present study and are revealed in
Fig.~\ref{figure:Hillel}.
\begin{enumerate}
 \item \emph{{Mixing.}} The vortices created during the inflation of
the bubble mix hot bubble gas with the ICM gas. This can be
clearly seen by following the evolution of the tracer. This is the
main heating mechanism of the ICM by jets \citep{GilkisSoker2012,
HillelSoker2014, HillelSoker2016}.
 \item \emph{Outward then inward motion.} The shocks running through the ICM
followed by sound waves behind them can push gas out. The traced
gas presented here is pushed from a torus cross section centered
at $(y,z)=(20,15)\kpc$ to an elongated shape centered at $(y,z)
\simeq (29,15)\kpc$ at $t \simeq 100 \Myr$. However, then it turns
around and flows inward. It is heated only when it starts to be
mixed with the hot bubble gas at $t \simeq 110 \Myr$
\citep{HillelSoker2016}.
 \item \emph{Large scale meridional flow.} At late times of $t \ga 200
\Myr$, and after about 10 jet-activity episodes (cycles), a
large-scale inflow takes place near the equatorial plane. This
results from the dragging of ICM gas by the outflowing jets along
their axis, here the horizontal $z$-axis. It seems though that in
reality the jets' axis will change direction and no such large
meridional flow will be develop. This change in direction can be
seen in the three bubble-pairs in NGC~5813
\citep{Randalletal2015}.
\end{enumerate}

{{{{ It should be noted that the simulations presented here are general, and have parameters that fit clusters of galaxies, and do not fit groups of galaxies, such as NGC~5813. For example, the power of jets here is 200-1000 times the power inferred for the bubble formation in NGC~5813 \citep{Randalletal2015}. The consequence is that the mixing region extends much beyond a radius of $30 \kpc$. For a much lower power of the jets as appropriate for NGC 5813, we would get a much smaller mixing region, one that fits the cooling flow region of NGC 5813. }}}}

The main conclusion of this section is that one cannot assume that
jets activity repeats itself along the same direction for many
episodes, while the ICM in the equatorial plane (the plane
perpendicular to the jets' axis) does not flow inward. The flow
inward will mix ICM gas with bubble gas, leading to heating by
mixing that is much more efficient than the shocks-heating
process.

\section{POST-SHOCK FLOW}
\label{sec:post}

\cite{Randalletal2015} find the observed temperature rise across
the shocks to be less than what is expected based on the Mach
numbers they derive from their calculated density jumps across the
shocks. They attribute the discrepancy to inability of the
deprojected temperature measurements to resolve the temperature
jump due to the narrow width of the shock and the rarefied cool
gas behind the shock \citep{Randalletal2011}.

\begin{figure}
\centering
\subfigure{\includegraphics[width=0.45\textwidth]{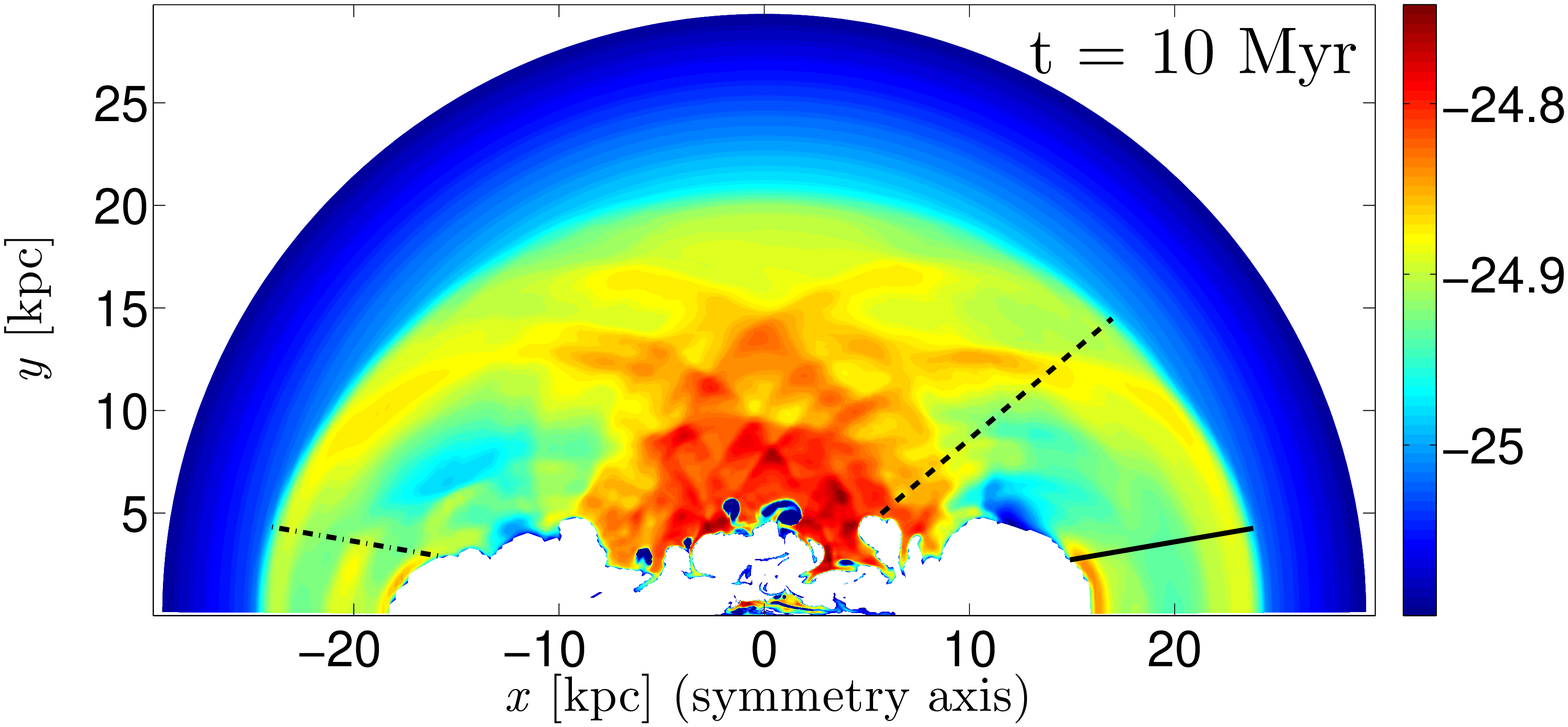}} \\
\subfigure{\includegraphics[width=0.45\textwidth]{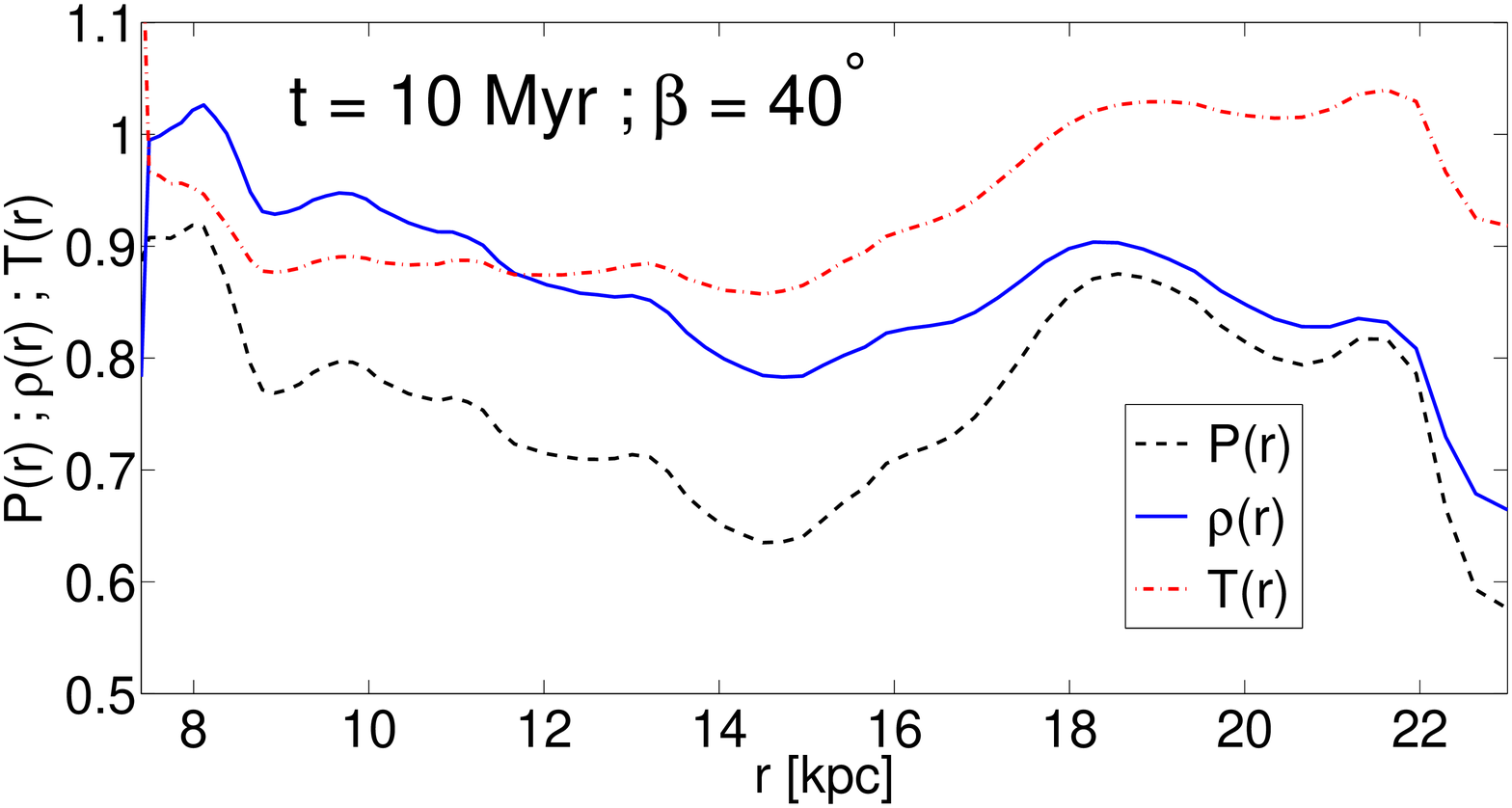}}
\vskip 1.0cm
\caption{Excitation of sound waves by wide-jet-inflated bubbles.
Upper panel: Density map for the ICM around the wide-jet-inflated
bubbles at $t=10 \Myr$ after the jets has ceased. The density
units of the color-code bar are $\log \rho(\g \cm^{-3})$. The
white regions are the two very-low density bubbles inflated by the
two opposite jets. Lower panel: Density, temperature and pressure
along the cut at $40^\circ$ as marked on the upper panel by the
dashed line. The pressure is given in units of $6 \times 10^{-10}
\erg \cm^{-3}$, density in units of $1.5  \times 10^{-25} \g
\cm^{-3}$, and temperature in units of $2.7 \times 10^7 \K$. Both
panels are taken from Sternberg \& Soker (2009) where all details
are given. The post-shock region at $r=21.5 \kpc$ shows jumps in
density, temperature, and pressure. The sound wave just behind the
shock at $r=18.5 \kpc$ has density and pressure above the
post-shock values, while the temperature is lower than the
post-shock value. }
  \label{figure:cutwide}
\end{figure}

\begin{figure}
\centering
{\includegraphics[width=0.45\textwidth]{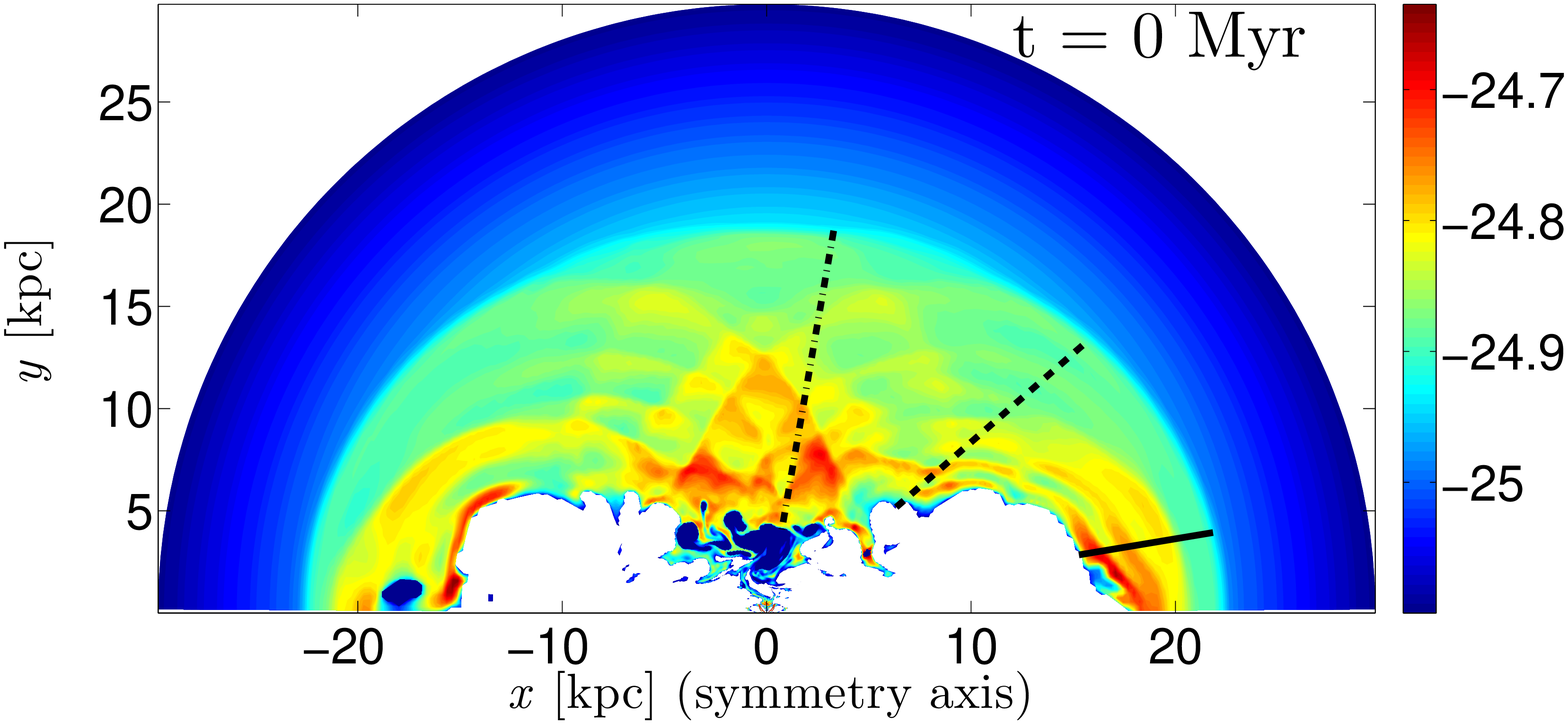}} \\
{\includegraphics[width=0.45\textwidth]{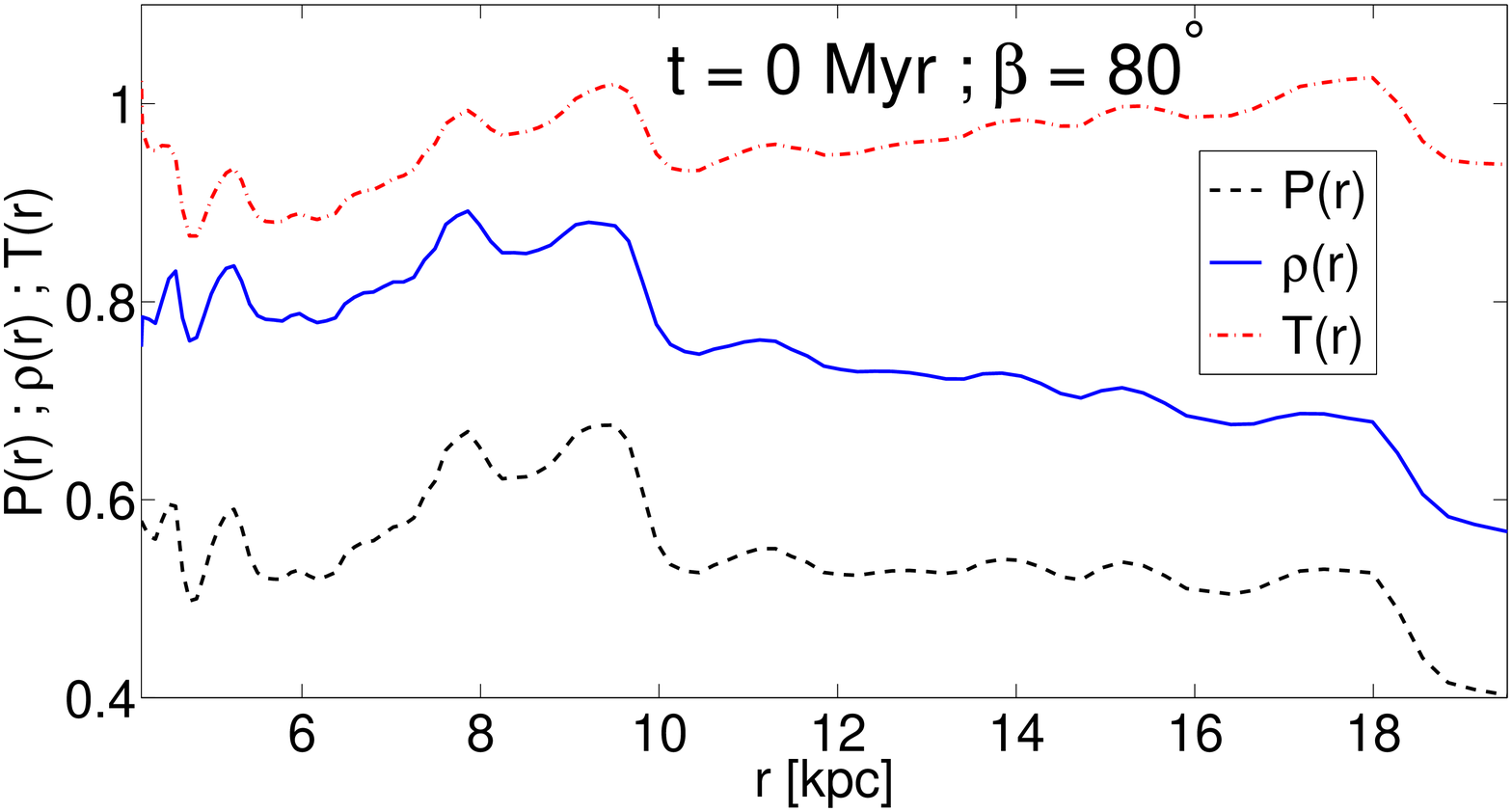}}
\vskip 1.0cm
\caption{Excitation of sound waves by precessing-jet-inflated
bubbles as simulated by Sternberg \& Soker (2009). Upper panel:
Density map for the ICM around the precessing-jet-inflated bubbles
at $t=0$, just when the jets cease. Lower panel: Density,
temperature and pressure along the cut at $80^\circ$ as marked on
the upper panel by the dashed-dotted line. The pressure is given
in units of $9 \times 10^{-9} \erg \cm^{-3}$, density in units of
$2  \times 10^{-25} \g \cm^{-3}$, and temperature in units of $2.7
\times 10^7 \K$. The post-shock region at $r=18 \kpc$ shows jumps
in density, temperature, and pressure. The sound wave behind the
shock at $r=15.5 \kpc$ has density and pressure above the
post-shock values, while the temperature is lower than the
post-shock value. }
  \label{figure:cutprecess}
\end{figure}

We here raise another possibility. When a bubble is inflated
by a jet sound waves are excited and trail the shock
wave  \citep{SternbergSoker2009}. The sound waves can compress gas
behind the shock, and raise the density behind the shock much more
than they raise the temperature. In Figs. \ref{figure:cutwide} and
\ref{figure:cutprecess} we present results from
\cite{SternbergSoker2009}, where all technical details are given.

The relevant properties of these simulations are as follows. The
simulation are performed in half of the meridional plane using the
two-dimensional version of the VH1 code \citep{Blondinetal1990,
Stevensetal1992} in spherical coordinates with a resolution of
$256 \times 256$. Gravity was included in these simulations. Two
opposite jets are launched along the $x$-axis.

In the wide jets case two opposite jets were injected in the time
period from $t=-10 \Myr$ until $t=0$, each jet with a half opening
angle of $\alpha = 70^\circ$. Their combined mass outflow rate was
$\dot M_{2{\rm j}}= 10 M_\odot \yr^{-1}$ and the initial velocity
was radial with $v_j= 7750 \km \s^{-1}$. The total power of the
two jets was $\dot E_{2{\rm j}}= 2\times10^{44} \erg \s^{-1}$.
This case is presented in Fig.~\ref{figure:cutwide}.

In the case of the precessing jets the two opposite jets have the
same mass outflow rate and velocity as in the wide-jets case, but
they were injected at any given time with a half opening angle of
$\alpha=5^\circ$. In the axisymmetric grid employed, the jets are
actually precessing very rapidly around the symmetry axis (i.e.,
in 3D torii were actually injected). The angle between the
symmetry axis and the jets' axis $\theta$ was varied in a random
way. The precession period, i.e., the time the jet returns to the
same angle $\theta$, is $T_{\rm prec}=0.1 \Myr$. The jet's
interaction with the ICM is similar to that of a wide jet with a
half opening angle of $\alpha \approx 50^\circ$. The jets were
active between $t=-18\Myr$ and $t=0$. This case is presented in
Fig.~\ref{figure:cutprecess}

The profiles along a cut in each case presented above show a
density and temperature jump immediately behind the shock, the
right edge in each of the two panels of profiles. Immediately
behind the post-shock jump (moving to the left in the panels)
there is another increase in the density, but the temperature
decreases. The over all density jump \emph{overestimates} the
shock Mach number.

We argue that \cite{Randalletal2015} somewhat overestimate the
Mach numbers in the shocks they study in NGC~5813.

\section{SUMMARY}
\label{s-summary}

We reexamined the claim made by \cite{Randalletal2011} and
\cite{Randalletal2015} that the ICM in the cooling flow group of
galaxies NGC~5813 is heated by many shock waves excited by
repeated jet-launching episodes. It was already argued by
\cite{Sokeretal2013} that the shocks-heating mechanism is
inefficient, and cannot account for the NGC~5813 ICM properties.
We here, however, performed three new analyses and confronted
theoretical studies with the newly derived properties of the ICM
of NGC~5813. ($i$) We presented a new analytical analysis of the
effect of many shocks on the outskirts of the ICM in NGC~5812.
($ii$) We discussed the role of the large scale flow induced by
jet-inflated bubbles on the ICM and its thermal evolution. ($iii$)
We used simulations of sound waves trailing shocks induced by
jet-inflated bubbles to deduce about possible complications in
deriving the shocks' properties from observations. These new
analyses strengthen the conclusion that mixing-heating is more
efficient that shock heating.

The energy carried by the shocks and that is not deposited in the
inner region, $r \la 30 \kpc$, is carried to the outer ICM regions
and must dissipate there as sound waves. In section \ref{sec:weak}
we found that for the inefficient shocks-heating mechanism to heat
the inner $r \la 30 \kpc$ region, this extra energy heats the
outer ICM to high enough temperature for it to escape the group of
galaxies. If this was the case, the inner region would have
expanded and escaped a long time ago. Clearly a more efficient
heating mechanism is required.

In section \ref{sec:flow} we showed that after about ten
jet-launching episodes along a constant axis a large-scale
meridional flow develops (Fig.\ref{figure:Hillel}). An inflow of
mass near the equatorial plane takes place. One cannot regard the
ICM for such a case as static. Such a flow leads to a very
vigorous mixing of the ICM with the hot bubbles' gas. Heating by
mixing becomes the dominate heating mechanism even for gas
perpendicular to the jets' axis \citep{GilkisSoker2012,
HillelSoker2014, HillelSoker2016}. We expect that in reality the
jets' axis will change over time, and the meridional flow will not
be as prominent as found here

In section \ref{sec:post} we showed that sound waves propagating
behind a shock wave can form a region behind the shock where the
density increases above the post-shock value, but the temperature
does not. This might explain the finding of \cite{Randalletal2015}
that the temperature jumps across the shocks in NGC~5813 are lower
than those expected for the Mach numbers they deduced based on the
density jumps. If true, it implies that \cite{Randalletal2015}
overestimated the Mach numbers in the three shocks they have found
in NGC~5813.

Our main conclusion is that shocks cannot be the main heating
process of the ICM in NGC~5813. Heating by mixing seems to be the
main heating process of the ICM by jets \citep{GilkisSoker2012,
HillelSoker2014, HillelSoker2016}. The mixing is caused by
vortices that are formed during the inflation of the bubbles. To
reveal the properties of the vortices in numerical simulations it
is crucial to inflate bubbles in a self-consistent manner. When
that is done, it is found that the mixing operates also
perpendicular to the jets' axis.

The vortices lead to turbulence in the ICM. Therefore, the finding
of moderate turbulence in the inner region of NGC~5813 by
\cite{Werneretal2009} and \cite{dePlaaetal2012}, as well as in
other clusters \citep{Zhuravlevaetal2014, Zhuravlevaetal2015,
AndersonSunyaev2015}, strengthens the argument that the
mixing-heating process is the dominate heating process. Some
heating by the dissipation of turbulence takes place as well, but
it accounts for at most $\approx 20\%$ of the total heating
\citep{HillelSoker2016}.

\section*{Acknowledgements}
{{{{ We thank an anonymous referee for helpful suggestions. }}}}


\begin{thebibliography}{}

\bibitem[Anderson \& Sunyaev(2015)]{AndersonSunyaev2015}  Anderson, M.,~E., \&
Sunyaev, R.\ 2015, arXiv:1506.01703

\bibitem[Arav et al.(2013)]{Aravetal2013} Arav, N., Borguet, B., Chamberlain, C., Edmonds, D., \& Danforth, C.\ 2013, \mnras, 436, 3286

\bibitem[Banerjee \& Sharma(2014)]{BanerjeeSharma2014} Banerjee, N. \& Sharma, P.\ 2014, \mnras, 443, 687

\bibitem[Blondin et al.(1990)]{Blondinetal1990} Blondin, J.~M.,
Kallman, T.~R., Fryxell, B.~A., \& Taam, R.~E.\ 1990, \apj, 356,
591

\bibitem[Cavagnolo et al.(2011)]{Cavagnolo2011} Cavagnolo, K.~W., McNamara, B.~R., Wise, M.~W., Nulsen,  P.~E.~J., Br{\"u}ggen, M., Gitti, M., \& Rafferty, D.~A.\ 2011, \apj, 732, 71

\bibitem[de Plaa et al.(2012)]{dePlaaetal2012} de Plaa, J., Zhuravleva, I.,
Werner, N., Kaastra, J. S., Churazov, E., Smith, R. K., Raassen,
A. J. J., \& Grange, Y. G.\ 2012, \aap, 539, AA34

\bibitem[Dunn et al.(2010)]{Dunn2010} Dunn, J.~P., Bautista, M.,  Arav, N., et al.\ 2010, \apj, 709, 611  

\bibitem[Edge et al.(2010)]{Edge2010} Edge, A.~C., et al.\ 2010, \aap, 518,
L47

\bibitem[Falceta-Goncalves et al.(2010)]{Falceta-Goncalves2010} %
Falceta-Goncalves, D., Caproni, A., Abraham, Z., Teixeira, D. M., \& de
Gouveia Dal Pino, E. M. 2010, ApJ, 713, L74

\bibitem[Farage et al.(2012)]{Farage2012} Farage, C.~L., McGregor, P.~J., \& Dopita, M.~A.\ 2012, \apj, 747, 28

\bibitem[Fogarty et al.(2015)]{Fogartyetal2015}  Fogarty, K., Postman,
 M., Connor, T., Donahue, M., \& Moustakas, J.\ 2015, \apj, 813, 117

\bibitem[Gaspari et al.(2012a)]{Gaspari2012a} Gaspari, M., Brighenti, F., \& Temi, P.\ 2012a, \mnras, 424, 190

\bibitem[Gaspari et al.(2012b)]{Gaspari2012b} Gaspari, M., Ruszkowski, M., \& Sharma, P.\ 2012b, \apj, 746, 94

\bibitem[Gilkis \& Soker(2012)]{GilkisSoker2012} Gilkis, A., \& Soker, N.\ 2012, \mnras, 427, 1482

\bibitem[Heinz \& Churazov(2005)]{Heinz2005} Heinz, S., \& Churazov, E.\ 2005, \apjl, 634, L141

\bibitem[Hillel \& Soker(2014)]{HillelSoker2014} Hillel, S., \& Soker, N.\ 2014, \mnras, 445, 4161

\bibitem[Hillel \& Soker(2016)]{HillelSoker2016}  Hillel, S., \& Soker, N.\ 2016, \mnras, 455, 2139

\bibitem[Li et al.(2015)]{Lietal2015} Li, Y., Bryan, G.~L., Ruszkowski, M., Voit, G.~M., O'Shea, B.~W., \& Donahue, M.\ 2015, \apj, 811, 73

\bibitem[McCourt et al.(2012)]{McCourt2012} McCourt, M., Sharma, P., Quataert, E., \& Parrish, I.~J.\ 2012, \mnras, 419, 3319

\bibitem[McNamara et al.(2014)]{McNamaraetal2014} McNamara, B.~R., Russell, H.~R., Nulsen, P.~E.~J., et al.\ 2014, \apj, 785,
44

\bibitem[Mignone et al.(2007)]{Mignone2007} Mignone, A., Bodo, G., Massaglia, S., et al.\ 2007, \apjs, 170, 228

\bibitem[Moe et al.(2009)]{Moe2009} Moe, M., Arav, N., Bautista, M.~A., \& Korista, K.~T.\ 2009, \apj, 706, 525

\bibitem[Mulchaey et al.(1996)]{Mulchaeyetal1996} {{{{ Mulchaey, J.~S.,
Davis, D.~S., Mushotzky, R.~F., \& Burstein, D.\ 1996, \apj, 456, 80 }}}}

\bibitem[Nesvadba et al.(2011)]{Nesvadba2011} Nesvadba, N.~P.~H., Boulanger, F., Lehnert, M.~D., Guillard, P., \& Salome, P.\ 2011, \aap, 536, L5

\bibitem[Omma et al.(2004)]{Omma2004} Omma, H., Binney, J., Bryan, G., \&
Slyz, A.\ 2004, \mnras, 348, 1105

\bibitem[Perucho et al.(2014)]{Peruchoetal2014} Perucho, M.,
Mart{\'{\i}}, J.-M., Quilis, V., \& Ricciardelli, E.\ 2014,
\mnras, 445, 1462

\bibitem[Pfrommer(2013)]{Pfrommer2013} Pfrommer, C.\ 2013, \apj, 779, 10

\bibitem[Pizzolato \& Soker(2005)]{Pizzolato2005} Pizzolato, F. \& Soker, N. 2005 \apj, 632, 821

\bibitem[Pizzolato \& Soker(2010)]{Pizzolato2010} Pizzolato, F., \& Soker, N.\ 2010, \mnras, 408, 961

\bibitem[Pope(2009)]{Pope2009} Pope, E.~C.~D.\ 2009, \mnras, 395, 2317

\bibitem[Prasad et al.(2015)]{Prasadetal2015} Prasad, D., Sharma, P., \& Babul A.\ 2015, \apj, 811, 10

\bibitem[Randall et al.(2011)]{Randalletal2011} Randall, S.~W., Forman,
W.~R., Giacintucci, S., et al.\ 2011, \apj, 726, 86

\bibitem[Randall et al.(2015)]{Randalletal2015} Randall, S.~W., Nulsen,
P.~E.~J., Jones, C., et al.\ 2015, \apj, 805, 112

\bibitem[Revaz et al.(2008)]{Revaz2008} Revaz, Y., Combes, F., \& Salom{\'e}, P.\ 2008, \aap, 477, L33

\bibitem[Roediger et al.(2007)]{Roediger2007} Roediger, E., Br{\"u}ggen, M.,
Rebusco, P., B{\"o}hringer, H., \& Churazov, E.\ 2007, \mnras, 375, 15

\bibitem[Russell et al.(2015)]{Russelletal2015} Russell, H.~R., Fabian, A.~C., McNamara,  B.~R., \& Broderick,
A.~E.\ 2015, MNRAS, in press

\bibitem[Sharma et al.(2012)]{Sharma2012} Sharma, P., McCourt, M., Quataert, E., \& Parrish, I.~J.\ 2012, \mnras, 420, 3174

\bibitem[Soker et al.(2013)]{Sokeretal2013} Soker, N., Akashi, M., Gilkis, A., Hillel, S., Papish, O., Refaelovich, M., \& Tsebrenko, D.\ 2013, Astronomische Nachrichten, 334, 402

\bibitem[Sternberg et al. (2007)]{Sternberg2007} Sternberg, A., Pizzolato, F. \& Soker N. 2007, \apj, 656, L5

\bibitem[Sternberg \& Soker(2008a)]{Sternberg2008a} Sternberg, A., \& Soker
N. 2008a, \mnras, 384, 1327 

\bibitem[Sternberg \& Soker(2008b)]{Sternberg2008b} Sternberg, A., \& Soker,
N.\ 2008b, \mnras, 389, L13 

\bibitem[Sternberg \& Soker(2009)]{SternbergSoker2009} Sternberg, A., \& Soker, N.\
2009, \mnras, 395, 228

\bibitem[Stevens et al.(1992)]{Stevensetal1992} Stevens, I.~R.,
Blondin, J.~M., \& Pollock, A.~M.~T.\ 1992, \apj, 386, 265

\bibitem[Sutherland \& Dopita(1993)]{SutherlandDopita1993} Sutherland, R.~S., \& Dopita, M.~A.\ 1993, \apjs, 88, 253

\bibitem[Tremblay et al.(2015)]{Tremblayetal2015}  Tremblay, G.~R., O'Dea, C.~P., Baum, S.~A. et al.\ 2015,
\mnras

\bibitem[Vernaleo \& Reynolds(2006)]{VernaleoReynolds2006} Vernaleo, J.~C., \& Reynolds, C.~S.\ 2006, \apj, 645, 83

\bibitem[Voit \& Donahue(2015)]{VoitDonahue2015} Voit, G.~M., \& Donahue, M.\ 2015, \apjl, 799, LL1

\bibitem[Voit et al.(2015)]{Voitetal2015} Voit, G.~M., Donahue, M.,
Bryan, G.~L., \& McDonald, M.\ 2015, \nat, 519, 203

\bibitem[Wagh et al.(2014)]{Waghetal2014} Wagh, B., Sharma, P., \& McCourt, M.\ 2014, \mnras, 439, 2822

\bibitem[Walg et al.(2013)]{Walgetal2013} Walg, S., Achterberg, A.,
Markoff, S., Keppens, R., \& Meliani, Z.\ 2013, \mnras, 433, 1453

\bibitem[Werner et al.(2009)]{Werneretal2009} Werner, N., Zhuravleva,
I., Churazov, E., Simionescu, A., Allen, S. W., Forman, W., Jones,
C., \& Kaastra, J. S.\ 2009, \mnras, 398, 23

\bibitem[Wilman et al.(2011)]{Wilman2011} Wilman, R.~J., Edge, A.~C., McGregor, P.~J., \& McNamara, B.~R.\ 2011, \mnras, 416, 2060

\bibitem[Zhuravleva et al.(2015)]{Zhuravlevaetal2015} Zhuravleva, I.,
Churazov, E., Arevalo, P., et al.\ 2015,  \mnras, 450, 4184

\bibitem[Zhuravleva et al.(2014)]{Zhuravlevaetal2014} Zhuravleva, I.,
Churazov, E., Schekochihin, A.~A., et al.\ 2014, \nat, 515, 85


\end{thebibliography}
\end{document}